\documentclass[useAMS,usenatbib]{mn2e}

\usepackage{psfig, epsf, epsfig}

\title[Discrete multiple populations in GCs]{The origin of discrete multiple stellar populations
in  globular clusters}
\author[K. Bekki,  T. Je\v{r}\'abkov\'a, P. Kroupa]
       {K. Bekki,${}^1$,
        T. Je\v{r}\'abkov\'a${}^{2,3}$,
        and P. Kroupa${}^{2,3}$\\
${}^1$ICRAR M468
The University of Western Australia
35 Stirling Hwy, Crawley
Western Australia 6009, Australia \\
        ${}^2$Astronomical Institute, Charles University in Prague, 
V Hole\v{s}ovi\v{c}k\'ach 2, CZ, 180-180 00 Praha 8, Czech Republic\\
        ${}^3$Helmholz Institut fur Strahlen Kernphysik, Universitat Bonn, 
Nussallee 14-16, 53115 Bonn, Germany}

\begin{document}

\date{Accepted, Received 2005 February 20; in original form }

\pagerange{\pageref{firstpage}--\pageref{lastpage}} \pubyear{2005}

\maketitle

\label{firstpage}

\begin{abstract}

Recent observations have revealed that at least several 
old  globular clusters
(GCs) in the Galaxy have discrete distributions of stars 
along the Mg-Al anti-correlation.
In order to discuss this recent observation,
we construct a new one-zone GC formation model
in which the maximum stellar mass ($m_{\rm max}$)
in the initial mass function (IMF) of stars in a forming GC depends  on
the star formation rate  (SFR), as deduced from independent observations.
We investigate the star formation histories of forming GCs.
The principal results are as follows. About 30 Myr after the formation of the
first generation
(1G) of stars within a particular GC,
new stars can be formed from ejecta from asymptotic
giant branch (AGB) stars of 1G. However, the formation of this second
generation (2G) of  stars can last only for [10-20] Myr,
because the most massive
SNe of 2G expel all of the remaining gas. The third generation (3G) of stars
are then formed from AGB ejecta $\approx 30$ Myr after the truncation of 2G
star formation. This cycle of star formation followed by its truncation
by SNe can continue until all AGB ejecta is removed from
the GC by some physical process.
Thus, it is inevitable that GCs have discrete multiple 
stellar populations
in the [Mg/Fe]-[Al/Fe] diagram. 
Our model predicts that  low-mass GCs are unlikely to have discrete
multiple stellar populations,
and  young massive clusters 
may  not have massive OB stars owing to low  $m_{\rm max}$
($<[20-30] {\rm M}_{\odot}$) during the secondary star formation.

\end{abstract}


\section{Introduction}

Old globular clusters (GCs) in the Galaxy are observed to have
anti-correlations between chemical abundances
of light elements (e.g., Carretta et al. 2009, C09; Cratton et al. 2012;
Renzini et al. 2015).
Since such anti-correlations are seen 
in almost all of GCs that have been investigated so far,
they are now considered to be essential characteristics of GCs.
Stars with  highly enhanced [Na/Fe]
(or [Al/Fe]) and severely depleted 
low [O/Fe]   (or [Mg/Fe])
along the Na-O anti-correlation
are often assumed to be formed from gas ejected from
previous generations of stars,  whereas those with 
[Na/Fe] and [O/Fe] similar to those of  the Galactic halo stars
are assumed to be formed from pristine gas within GC-forming
molecular  clouds.
The origin of the anti-correlations 
has been discussed by several authors
in the context of GC formation processes
(e.g., Fenner et al. 2004; Bekki et al. 2007, B07;
Prantzo \& Charbonnel 2006, PC06;
D'Ercole et al. 2010, D10, Ventura et al. 2016; Bekki 2017a, B17a).
For example, 
the number fraction of "polluted" stars with high [Na/Fe] ([Al/Fe])
and low [O/Fe] ([Mg/Fe]) in a GC  has been used to provide fossil information
on the original  mass the GC and the initial mass function
of stars in the early phase of the GC formation
(e.g., Smith \& Norris 1982; D'Antona \& Caloi 2004;
Bekki \& Norris 2006; PC06).
Previous theoretical studies of GC formation tried
to explain the apparently continuous distribution of stars
along the anti-correlation between light element by assuming
dilution of AGB ejecta with pristine gas
(e.g., B07; D10).

Although the distributions of stars along the anti-correlations
were assumed  to be continuous in previous observational and theoretical
studies of GCs,
more precise spectroscopic measurements of chemical abundances
of GC stars by
Carretta (2014; C14) have recently revealed that (i) the distribution of stars
along the [Mg/Fe]-[Al/Fe] anti-correlation in NGC 2808 is not
continuous and (ii) there are three distinct groups each of which
has different [Mg/Fe] and [Al/Fe].
Such discrete three stellar populations have been also found in 
NGC 6752 (Carretta et al. 2012; Milone et al 2013), though
they are  less clear in comparison with NGC 2808.
Marino et al. (2011) also found two distinct groups each of
which clearly shows the Na-O anti-correlation in
the [Na/Fe]-[O/Fe] diagram for M22.
If the presence of  discrete multiple stellar populations 
along the anti-correlations between light elements is a universal
feature in old GCs, then it can give a strong constraint on the theory
of GC formation.
The discrete stellar populations  in GCs would imply that 
such GCs experienced a number of discrete star formation episodes in their
early formation histories.

The origin of these  new observational results  on the discrete
multiple stellar populations of GCs, however,
have not been discussed
so extensively in theoretical studies of GC formation.
Renzini et al. (2015) pointed out that the observed discreteness
can not be explained by a GC formation scenario in which later generations
of stars are formed from ejecta from fast rotating massive stars (FRMS)
and massive interacting binaries.
Accordingly,  the discrete multiple populations in GCs  could possibly
give strong constrains on the formation processes of GCs.
Using new hydrodynamical simulations of GC formation from fractal molecular
clouds,
Bekki (2017b, B17b) have recently shown that  
forming GCs can have a number of bursty star formation events in their
early ($<300$ Myr) evolution. However,  such a result 
is due largely to the adopted model in which
ejection of gas from AGB stars can occur only at five separate times
(not continuously) for some  numerical reasons.
Therefore, it remains unclear how forming GCs can have discrete multiple
stellar populations.

D'Antona et al. (2016) discussed the origin of at least five discrete populations
in NGC 2808 by assuming (i) several distinct episodes of star formation, 
(ii) gas fueling from AGB stars with different masses (thus different yields),
and (iii) dilution of the AGB ejecta with pristine gas.
They have shown that the first population (25\% of the GC) enriched in 
N yet not so enriched in He and Na
can be formed from  gas from lower mass AGB stars diluted by pristine gas.
They also have demonstrated that NGC 2808 has a minor population that 
was chemically polluted by iron-rich ejecta from SNIa.
However, they did not discuss the physical basis for the assumed several episodes
of star formation in  a quantitative manner.
They did not provide the physical reasons for the assumed no SNII in 2G
populations of GCs either.
Thus, it is still theoretically unclear how discrete multiple stellar populations can be
formed in GCs.

If the IMF in secondary star formation from AGB ejecta is an invariant canonical one,
then the star formation can be  truncated by SNe
of massive stars only $3 \times 10^6$ yr after its commencement,
because such SN explosions
can easily expel all of the remaining gas within GCs 
(D'Ercole et al. 2008, D08; D10; B17b). 
Therefore, a canonical IMF can be a potentially
serious problem in the AGB scenario.  A way to avoid this problem
is to adopt a non-standard IMF in which the upper mass cut-off
($m_{\rm u}$)  of the IMF
is less than $9 {\rm M}_{\odot}$ (i.e., virtually no SNe) in
the formation of later generations of stars from AGB ejecta.  However,
it is totally unclear how such an apparently
unusual IMF is possible during GC formation.
Also no previous theoretical studies of GC formation have ever discussed
whether and how the observed discrete multiple stellar populations 
can be achieved in GC formation with a non-standard IMF.

The purpose of this paper is thus to investigate the origin of 
discrete multiple stellar 
populations in GCs in the context of a non-universal IMF, 
by assuming that the later 
stellar generations form in smaller clusters and star formation 
is interrupted by SN II in the later generations.
To describe clustered star formation,
we use empirically motivated 
physical and mathematical formalism of the IGIMF theory (Kroupa et al. 2013).
We emphasise that the IGIMF theory is applied here without 
adjustments for this GC case. The results achieved are 
therefore not fine-tuned to reach an aim.
We particularly  investigate
the star formation histories 
of GCs over $\approx 400$ Myr 
using a new GC formation model that incorporates a non-universal
IMF self-consistently.
A growing number of observational and theoretical
studies have recently discussed that the IMF in GC
formation can be non-universal using the observed different properties of
GCs (e.g., D'Antona \&  Caloi 2004; Bekki \& Norris 2006; PC06;
Marks et al. 2012).  
For example,
Marks et al. (2012) have recently estimated the IMF slope for high-mass stars
(i.e., $\alpha_3$ in the Kroupa IMF; Kroupa 2001) for each GC in the Galaxy
using the observed
mass density of  the GC and found that $\alpha_3$ was quite different
between different GCs.
Accordingly, 
it is quite important and timely for 
the present study to investigate GC  formation 
based on a non-universal IMF.

\begin{table}
\centering
\begin{minipage}{90mm}
\caption{Physical meanings of acronyms and model parameters.}
\begin{tabular}{ll}
{Acronym 
 } &
{Physical meaning
} \\
1G & First generation of stars formed from original gas \\
2G & Second generation of stars formed from ejecta  from 1G\\
3G & Third generation of stars formed from ejecta  from 1G\\
$n$G & $n$th generation of stars formed from ejecta from 1G \\
LG & Later generation of stars formed from AGB ejecta \\
$\Delta t_{\rm sf}$ & Duration of star formation (SF) episode \\
$\Delta t_{\rm tr}$ & Duration of SF truncation (no SF) \\
$m_{\rm max}$ & The maximum stellar mass in each SF episode \\
MC & Molecular cloud forming a GC \\
$M_{\rm mc}$ & The initial mass of a MC \\
$M_{\rm g}$ & The total  mass of gas \\
$M_{\rm 1G}$ & The total mass of 1G stars \\
$M_{\rm LG}$ & The total mass of LG stars \\
${\dot{M}}_{\rm inf}$ & The gas infall rate onto the core of a MC \\
${\dot{M}}_{\rm agb}$ & The gas ejection rate of AGB stars \\
$m_{\rm agb,l}$ & The lowest mass of AGB stars for LG formation \\
SFR & Star formation rate \\
IGIMF &  Integrated galactic initial mass function \\
\end{tabular}
\end{minipage}
\end{table}

The structure of the paper is as follows.
We describe the new one-zone  models of GC formation
in \S 2.
We present the results on the star formation histories of forming GCs
for different representative models 
in \S 3.
We  describe the possible spreads of [Mg/Fe] and [Al/Fe] of
GC stars based on the models in \S 4.
We also discuss the origin of the observed discrete multiple stellar populations
of the Galactic GCs based on the present results in \S 4.
We  provide important implications of the present results in terms of
the origin of multiple stellar populations in GCs in this section
in \S 4.
We summarize our  conclusions in \S 5.
Although several authors have recently discussed a number of new physical
processes related to the origin of multiple stellar populations
of GCs, such as merging of  GCs (Bekki \& Yong 2012; Bekki \& Tsujimoto 2016),
stripping of stellar envelopes in massive stars
(e.g., Elmegreen 2017),
gas accretion onto pre-main sequence stars (e.g., PC06),
and Bondi-accretion of interstellar medium  onto GCs (e.g.,
Pflamm-Altenburg \&  Kroupa 2009),
we do not discuss them extensively in the present study.
We do not discuss all of the AGB scenario (e.g., the yields, the mass-budget,
the dilution problems etc), instead, focus only on the discrete populations
in the present paper.

\begin{figure}
\psfig{file=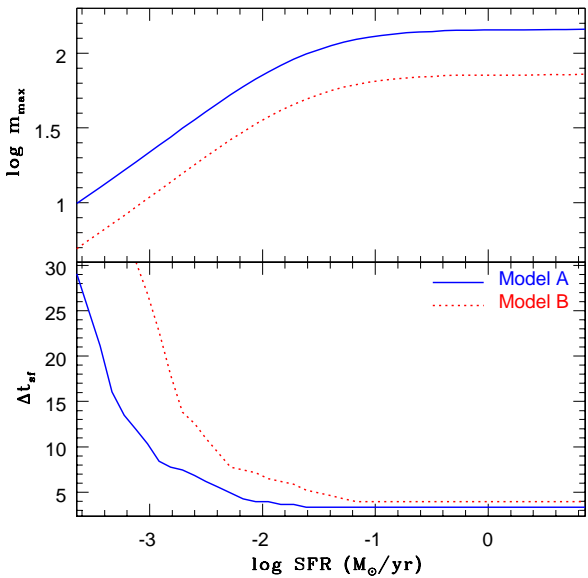,width=8.5cm}
\caption{
$SFR-m_{\rm max}$ (upper panel)
and $SFR-\Delta t_{\rm sf}$ relation (lower panel) derived from
two IMF models, model A (blue solid) and B (red dotted).
Here $m_{\rm max}$ and $\Delta t_{\rm sf}$
represent the maximum mass of stars 
and the duration of star formation in each star formation episode,
and it corresponds to the lifetime of the most massive star formed,
respectively.
The two relations in  model A  are derived from 
 the IGIMF theory recently developed by several authors,
e.g., Kroupa \& Weidner (2003),
Kroupa et al. (2013), Weidner et al. (2013), and Yan et al. (2017)
based on observational results.
In model B, the relations are 
a slightly modified version of the relations in model A:
$m_{\rm max}$  at a given
SFR in model B is by a factor of two smaller than that in model A.
The details of model A and B are given in the main text.
}
\label{Figure. 1}
\end{figure}

\begin{table}
\centering
\begin{minipage}{85mm}
\caption{Physical parameters for
the representative 11 models.}
\begin{tabular}{lllll}
{Model ID \footnote{ M1 is the model in which truncation
of star formation by SNe is not considered at all.
A canonical (fixed) IMF is adopted in M9,
$m_{\rm max}$ is also a constant ($100 {\rm M}_{\odot}$; fixed duration
of star formation in later generations of stars) 
The IGIMF theory is adopted for M2-M9 and M11 to calculate
$m_{\rm max}$ self-consistently.
The IMF slope in the formation of 1G stars is assumed to be
$\alpha=$1.35 (top-heavy) for the model M10.
The total mass of 1G in the GC of M11 is less than
$10^6 {\rm M}_{\odot}$ (less than 10\% of the original gas mass).
 }} & 
{$M_{\rm mc}$ \footnote{ The initial total mass of a GC-forming
molecular cloud (MC) in units of $10^7 {\rm M}_{\odot}$
}} & 
{$SFR-m_{\rm max}$ \footnote{ `A' and `B' represent the model name
for the physical relation between SFR (star formation rate) and
$m_{\rm max}$ (the maximum mass of stars in the adopted variable IMF).
The details of models A and B are given in the main text and Fig. 1.
 }} & 
{$m_{\rm  agb,l}$  \footnote{ The lower mass cut-off 
(${\rm M}_{\odot}$) of AGB stars above
which AGB ejecta is assumed to be converted into new stars. 
}} & 
{$C_{\rm sf}$ \footnote{ Star formation coefficient. See eq. (2).
 }}  \\
M1 & 1.0&  $-$  & 4.0 & 9.0 \\
M2 & 1.0&  A  & 4.0 & 9.0 \\
M3 & 1.0&  A  & 3.0 & 9.0 \\
M4 & 1.0&  A  & 3.0 & 18.0 \\
M5 & 1.0&  B  & 3.0 & 9.0 \\
M6 & 0.1&  A  & 3.0 & 9.0 \\
M7 & 0.3&  A  & 3.0 & 9.0 \\
M8 & 3.0&  A  & 3.0 & 9.0 \\
M9 & 1.0&  $-$ (fixed $m_{\rm max}$)  & 3.0 & 9.0 \\
M10 & 1.0&  A (top-heavy IMF)  & 3.0 & 9.0 \\
M11 & 1.0&  A   & 3.0 & 4.5 \\
\end{tabular}
\end{minipage}
\end{table}

\section{The GC formation model}

We here adopt the ``AGB pollution scenario'' in which later generations
(LG) of stars can be formed from gas ejected from AGB stars
having evolved from intermediate-mass  first generation
(1G) stars formed within a GC-forming molecular cloud (MC).
We use the term ``1G'' and ``LG'' rather than ``FG'' and ``SG''
to represent the first and later generations of stars of the GC in the present
study, because not just second or third
(2G and 3G, respectively), but fourth and fifth generations (4G and 5G)
of stars can be formed in the present GC formation models.
Table 1 describes the physical meaning of these acronyms used
in the present study.
Also we focus exclusively on this scenario, though other
GC formation scenarios based on chemical pollution by 
stars other than AGB stars (e.g., FRMSs;  Decressin et al. 2007)
are possible. These scenarios, however,
can not simply explain the discrete multiple stellar populations
(e.g., Renzini et al. 2015).

\begin{figure*}
\psfig{file=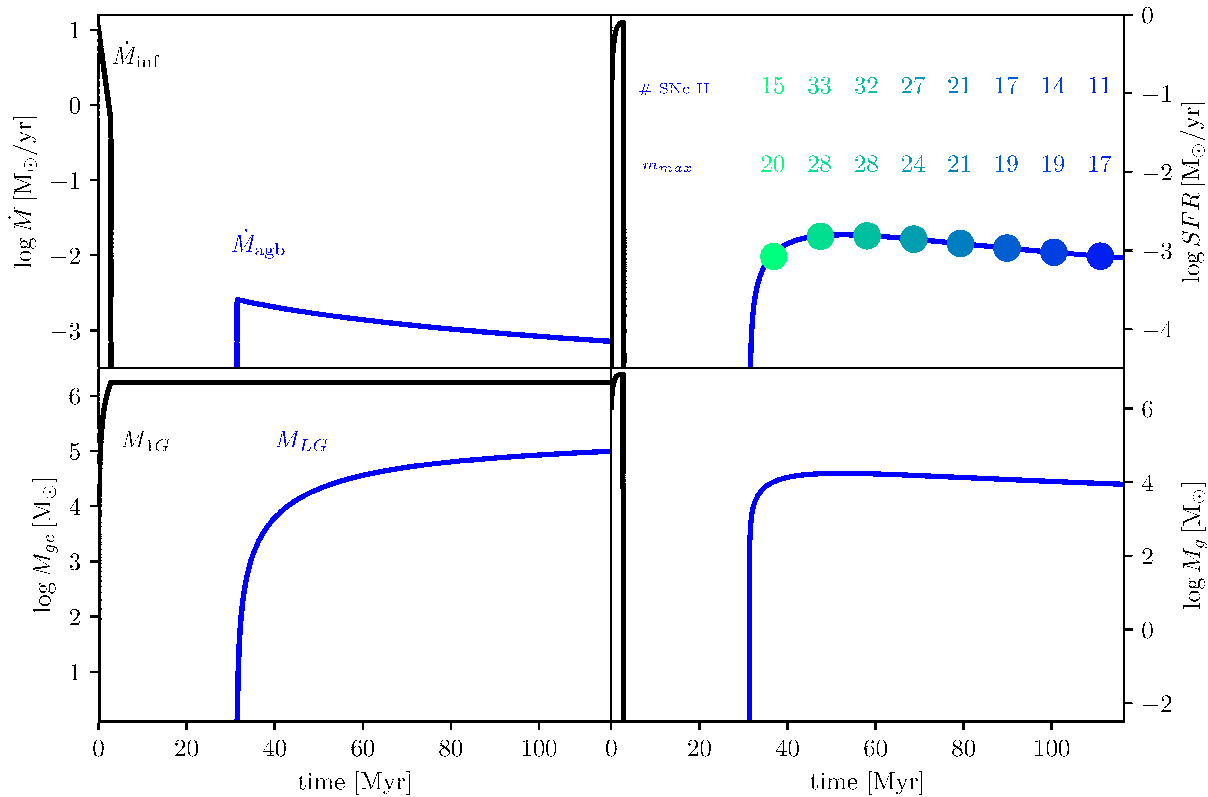,width=16.0cm}
\caption{
Time evolution of
gas accretion and AGB ejection rates (upper left),
SFR (upper right),  total masses of stars in 1G and LG (lower left),
and total gas mass (lower right) in the model M1 (See the details
of the model parameter in Table 2).
The total number of type II SNe ($N_{\rm SNII}$) and $m_{\rm max}$
in units of $M_{\odot}$  (maximum mass
of stars) in each of the selected time intervals are shown
in the upper right panel.
The end of 1G star formation at $T \approx 30$ Myr corresponds to
$T=0$ in Fig.3.
}
\label{Figure. 2}
\end{figure*}

\subsection{Basic equations in one-zone models}

We adopt ``one-zone''  models of GC formation (i.e., 1G and LG formation)
in which  
a GC is assumed to form through a continuous
gas infall onto the core of a GC-forming MC
with the initial total mass of $M_{\rm mc}$.
The basic equation for the evolution of gas and new stars is as follows:
\begin{equation}
\frac{dM_g}{dt}=-\psi(t)+A(t)+w(t),
\end{equation}
where  $M_{\rm g}$ is the total gas mass within the star-forming
core of the MC,
$\psi(t)$ is the star formation rate,
$A(t)$ is the rate of gas accretion onto the core of the MC,
and $w(t)$ is the injection rate of gas from its AGB stars.
The star formation rate $\psi(t)$ (SFR) is assumed to be proportional
to the gas mass ($M_{\rm g}$)  with a constant star formation
coefficient and thus is described as follows:
\begin{equation}
\psi(t)=C_{\rm sf} M_{\rm g}(t).
\end{equation}
This parameter  $C_{\rm sf}$ describes the rapidity
of star formation within a GC-forming MC.
Since the present one-zone model can not investigate 
the details of star formation processes
within GC-forming MCs, we need to adopt a reasonable value
for $C_{\rm sf}$.
Our recent hydrodynamical simulations of star formation in GC-forming
MCs with $M_{\rm mc}=10^7 {\rm M}_{\odot}$ shows 
that (i) the peak star formation for 
1G stars is about $1 {\rm M}_{\odot}$ yr$^{-1}$
and (ii) the peak star formation in LG stars is 
$[0.001-0.01] {\rm M}_{\odot}$ yr$^{-1}$ (see Fig. 6 in B17b). 
These are broadly consistent with the results of the fiducial model
with $C_{\rm sf}=9$, which means that the adopted
$C_{\rm sf}$ is reasonable and realistic.
Furthermore, the models with the adopted
$C_{\rm sf}=[9-18]$ demonstrate that the total mass of 
1G stars 
can be larger than  $\approx 10^6 {\rm M}_{\odot}$ for
$M_{\rm mc} \approx 10^7 {\rm M}_{\odot}$.
This means that an enough amount of gas can be ejected from AGB stars
for LG formation.
Since a  large fraction
of 1G stars can be lost in the later evolution of a GC
(e.g., Vesperini et al. 2011; Rossi et al 2016),
the initial mass of FG stars should be as large as 
$\approx 10^6 {\rm M}_{\odot}$ for a typical GCs with the present-day mass of 
$M_{\rm gc}= 2 \times 10^5 {\rm M}_{\odot}$ to be formed.

We assume that $C_{\rm sf}$ is the same for 1G and LG formation in the present
study.
The present models with a constant  $C_{\rm sf}$ is quite reasonable,
because they
can predict star formation histories of 1G and LG that are similar to
those derived by fully self-consistent hydrodynamical simulations of GC formation
(Bekki 2017b).
Also, it is not realistic for the present study to 
adopt a variable $C_{\rm sf}$,
because it is theoretically 
unclear how $C_{\rm sf}$ depends on physical properties of GC-forming
MCs.

Conroy \& Spergel (2011) suggested that the effects of Lyman-Werner (LW) photons 
from 1G stars on 
${\rm H_2}$ (in AGB ejecta)
are very important, because they can prevent secondary star formation owing to the
dissociation of ${\rm H_2}$. 
However, they did not consider the importance of ${\rm H_2}$ self-shielding
from the intense LW radiation. Such a self-shielding effect is very important
for gas with ${\rm H_2}$ column density ($N_{\rm H2}$)
significantly larger than $10^{14}$ cm$^{-2}$
(e.g., Draine \& Bertoldi 1996). The reduction factor ($f_{\rm red}$)
 can be approximated
described as follows:
\begin{equation}
f_{\rm red}=
{ ( \frac{ N_{\rm H_2}  }{ 10^{14} {\rm c}m^{-2}  } ) }^{-0.75} .
\end{equation}
A lower $f_{\rm red}$ means that a larger amount of LW photons 
can be self-shielded by ${\rm H_2}$ (See Draine \& Bertoldi 1996 for
a more complex functional form for $f_{\rm ret}$).
As shown in our recent simulations (B17b; Bekki \& Tsujimoto 2017),
the density of intra-cluster gas ejected 
from AGB stars ($\Sigma_{\rm icm}$)
 in a forming GC can be much higher than $10^{14}$ cm$^{-2}$
($\Sigma_{\rm H_2} >10^{22}$ cm$^{-2}$)
 within the central 1pc. This means that the suggested
suppression of star formation by the LW photons is completely negligible.
Thus, our assumption of LG formation soon after gas accumulation in the centers of GCs
is reasonable.

We consider that the formation of 1G stars can be truncated by
SNe $3 \times 10^6$ yr after the initial burst of star formation
in a GC-forming MC. This SF-duration ($\Delta t_{\rm sf}$)
corresponds to the lifetime of the most massive star
(with a mass of $\approx 100 {\rm M}_{\odot}$) within the MC.
We also consider two different models for star formation from 
AGB ejecta of 1G stars  in  a GC  as
follows.
One is that star formation can continue 
over $\approx 400$ Myr without any interruption of star formation
 by energetic events
such as supernova (SNe). 
(`continuous model').
The other is that star formation
can be  truncated by SNe from LG stars owing to the strong thermal
and kinematic feedback  (`multiple burst model', though,
literally, it is not a burst but a sporadic low-level star formation).
In the previous one-zone GC formation models
(e.g., B07; D08,D10),
star formation from AGB ejecta was assumed  to be continuous
for a certain timescale (of an order of $10^8$ yr). However, we consider
that such an assumption is highly unrealistic, given the lifetime
of massive stars with $m >10 {\rm M}_{\odot}$ is quite short
(less than $2 \times 10^7$ yr).
We here investigate mainly a number of
multiple burst models, though we investigate
the continuous models too
just for comparison.

In the multiple burst model, $C_{\rm sf}$ is set to be 0 when the 
most massive star with a mass of $m_{\rm max}$ within a 
generation of stars (e.g., 2G, 3G, and 4G) explodes as a SN.
All of the remaining gas (i.e., AGB ejecta) is assumed to be expelled
owing to the strong SN feedback effects after the SN (i.e., SFR=0).
Accordingly, the duration of star formation ($\Delta t_{\rm sf}$)
corresponds to the lifetime ($t_{\rm lf}$)  of the most massive star 
($t_{\rm lf} (m_{\rm max}) $).
Both $C_{\rm sf}$ and $M_{\rm g}$ are set to be 0 until all of the 
SNe are exploded, and this period of SF truncation is denoted
as $\Delta t_{\rm tr}$. We adopt a fixed $\Delta t_{\rm tr}$ of 
$3.2 \times 10^7$ yr for all models in the present study,
since it approximately corresponds to the $[8-9] {\rm M}_{\odot}$ SN.
Recent studies of the IMF
have demonstrated  that the IMF can be time-evolving and 
depend on a few key parameters of star-forming regions, such as
metallicity, SFR, and gas density (e.g., Marks et al. 2012; Kroupa
et al. 2013).
In order to estimate $m_{\rm max}$ in each episode of star formation,
we use the results of these theoretical studies on the correlation
between $SFR$ and $m_{\rm max}$ (e.g., Kroupa et al 2013; Yan et al. 2017;
Stephens et al. 2017).

We adopt 
a standard $m-t_{\rm lf}$ relation for stars with $m\ge 9 {\rm M}_{\odot}$
in order to derive 
the $SFR - \Delta t_{\rm sf}$ relation from
the $SFR-m_{\rm max}$ relation.
Fig. 1 shows the adopted two models for the $SFR - \Delta t_{\rm sf}$ 
relations. Model A is consistent with the theoretically derived
one by Yan et al. (2017) whereas $m_{\rm max}$ in   model B
is by a factor of 2 smaller
than that in model A. Therefore, $\Delta t_{\rm sf}$ is appreciably
longer in  model B than in  model A.
This model B could be also derived from the IGIMF 
theory for star clusters, 
if the mass function of very young clusters 
becomes bottom heavy at low SFRs.
We estimate
$\Delta t_{\rm sf}$ 
at each time step, $T=t_i$,  using 
the $SFR - \Delta t_{\rm sf}$ relation and compare 
$\Delta t_{\rm sf}$ 
with 
$t_i-t_{\rm start}$,
where $t_{\rm start}$ is the time at which star formation started.
If $\Delta t_{\rm sf}$ is shorter than $t_i-t_{\rm start}$,
then star formation is truncated from $T=t_{\rm i}$ to 
$T=t_{\rm i}+\Delta t_{\rm tr}$.

\begin{figure}
\psfig{file=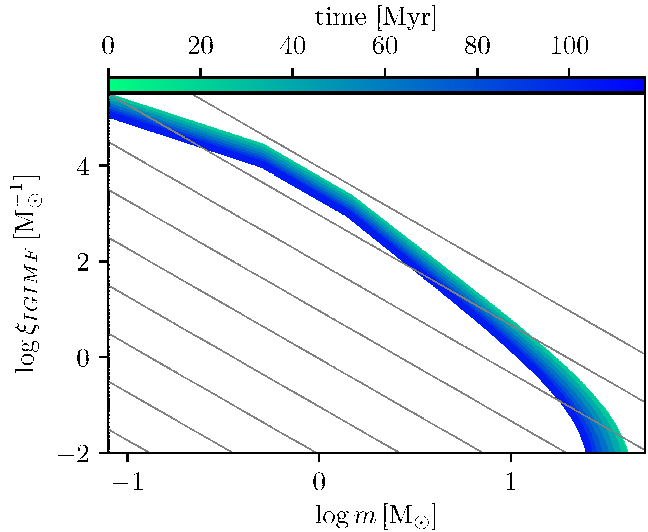,width=8.5cm}
\caption{
IGIMF as a function of stellar mass ($m$) at different epochs
in the model M1. At early times, the IGIMF is less top-light whereas it
is less top-heavy at later times (when SFR decreases with time).
The diagonal lines indicate the canonical IMF with $\alpha=2.35$.
}
\label{Figure. 3}
\end{figure}

\begin{figure*}
\psfig{file=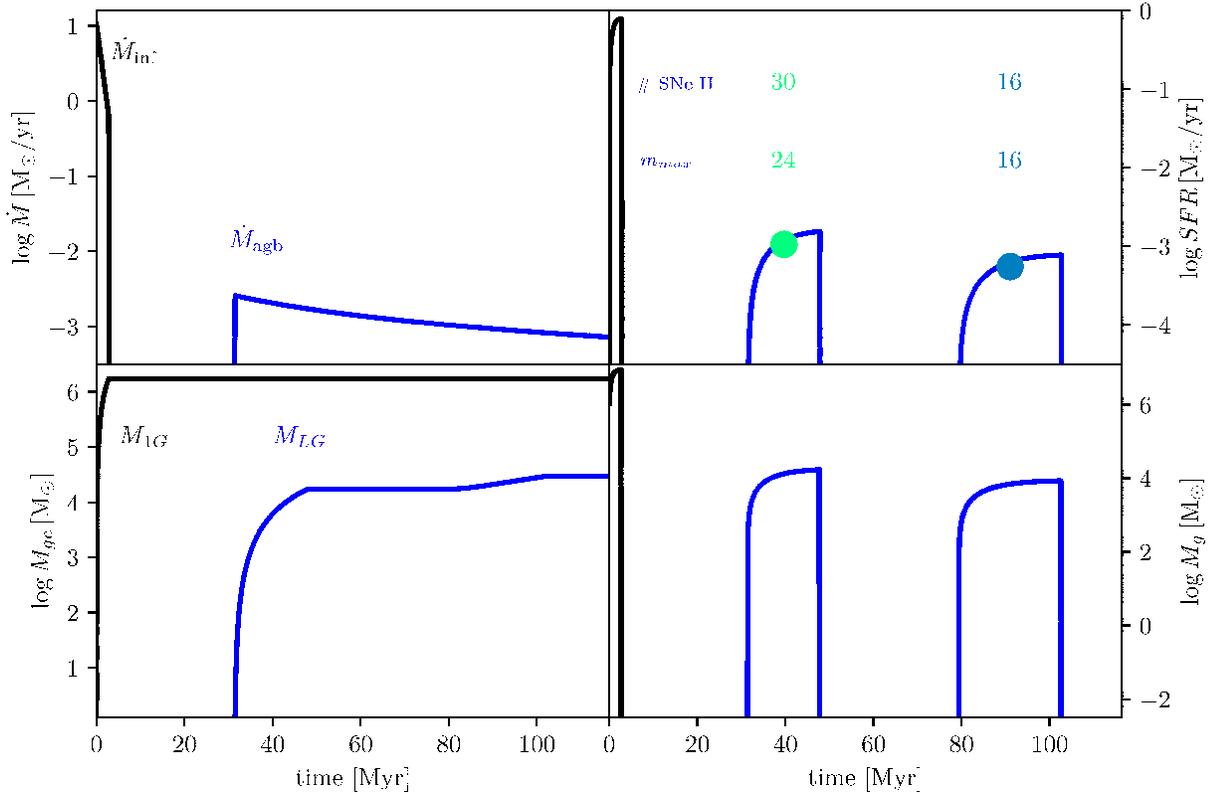,width=16.0cm}
\caption{
The same as Fig. 2 but for M2 in which the duration of star formation
is self-consistently derived from the evolution of the SFR.
}
\label{Figure. 4}
\end{figure*}

For the accretion rate, we adopt 
$A(t)=C_{\rm a}\exp(-t/t_{\rm a})$
and $t_{\rm a}$ is set to be $10^6$ yr.
We confirm that the present results do
not depend on $t_{\rm a}$ as long as it is shorter than
$3 \times 10^6$ yr.
The normalization factor $C_{\rm a}$ is determined such
that the total gas mass accreted from AGB ejecta onto the core of a
 MC  can be $M_{\rm mc}$ for a given
$t_{\rm a}$. 
In order to estimate the total mass of AGB ejecta in a  GC,
we adopt the IMF that is   defined
as $\Psi (m) = C_{0}m^{-\alpha}$,
where $m$ is the initial mass of
each individual star and the slope $\alpha =2.35$
corresponds to the canonical IMF  (Salpeter 1955; Kroupa 2001).
The normalization factor $C_0$ is a function of $\alpha$,
$m_{\rm l}$ (lower mass cut-off; $0.1 {\rm M}_{\odot}$), 
and  $m_{\rm u}$ (upper mass cut-off).
We investigate models with different $\alpha$ to discuss
how the total mass of ejecta from AGB stars depend on $\alpha$
and how it can influence the SFR of LG stars.
It should be noted here that although this single power-law IMF for 1G 
is not so realistic as those adopted in recent IMF studies 
(e.g., Marks et al. 2012; Kroupa et al. 2013),
this approximation is sufficient for the purpose of predicting
the total mass of AGB ejecta in this study.

The total mass of gas ejected from AGB stars between
$T=t$ and $T=t+\delta t$ ($M_{\rm agb}$),
where $\delta t$ is the time step width,
is  described as:
\begin{equation}
M_{\rm agb}=\int_{m_{\rm agb}(t+\delta t)}^{m_{\rm agb}(t)} m_{\rm ej}
\Psi (m_{\rm ini}) dm_{\rm ini},
\end{equation}
where $m_{\rm ej}$ describes the total gas mass ejected from
an  AGB star with initial mass  $m_{\rm ini}$
and final mass  ($m_{\rm fin}$).
The lowest ($m_{\rm agb}(t+\delta t)$) 
and highest masses ($m_{\rm agb}(t)$)
of AGB stars at $T=t$ correspond to the masses of stars which
enter into the main-sequence turn-off at $T=t+\delta t$ and $T=t$,
respectively,
We adopt the following  analytic form of $m_{\rm ej}$
derived by Bekki (2011) based on observational results by
(Weidemann 2000):
\begin{equation}
m_{\rm ej} =0.916 m_{\rm ini}-0.444,
\end{equation}
where $m_{\rm ej}$ and $m_{\rm ini}$ are given in units of ${\rm M}_{\odot}$.
Thus, the AGB wind rate is as follows:
\begin{equation}
w(t) =\frac{ dM_{\rm agb} }{ dt }
\end{equation}
In order to calculate $t_{\rm lf}$
of stars with $m \le 9 {\rm M}_{\odot}$,
we use the mass-age relation by
Renzini \& Buzzoni (1986; 2010).

\subsection{IGIMF}

\begin{figure}
\psfig{file=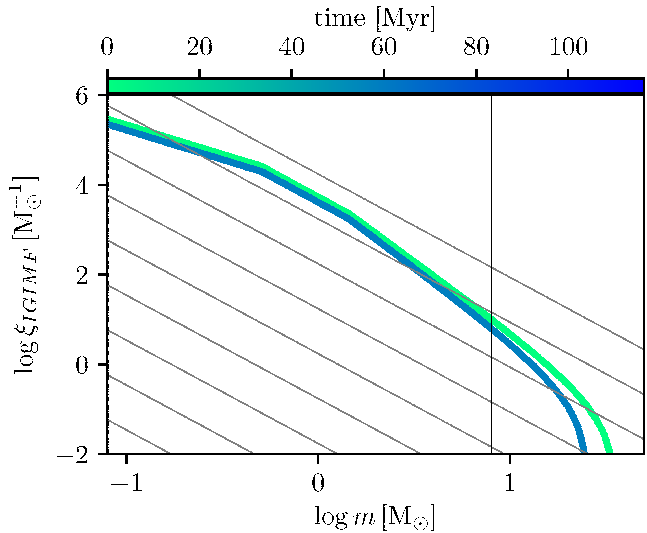,width=8.5cm}
\caption{
The same as Fig. 3 but for M2.
}
\label{Figure. 5}
\end{figure}

The standardly made assumption is that the 1G and LGs are 
formed with an invariant IMF.  This leads to
the problem that the LGs produce 
too many SN explosions which inhibit the 
build-up of the LGs such that the AGB scenario for 
LGs has been often thought to be less realistic.
Closer scrutiny of the observational data however 
indicates that star formation occurring in 
embedded clusters might result in a  non-standard IMF
(i.e., non-standard composite IMF of the whole stellar populations
in all clusters).
Even previously-thought distributed star formation has been found to be 
organized in embedded clusters (e.g. fig.12  in   Megeath et al. 2016), 
whereby low-intensity star formation produces low-mass embedded 
clusters (ECs) only. Low-mass ECs do not 
contain massive stars because the molecular gas mass is 
distributed over the stellar population in the form of a largely invariant 
IMF (Kroupa 2002; Bastian et al. 2010;
Marks et al. 2012; Kroupa et al. 2013) leaving not enough mass to form
 a massive star within the EC. Thus, for example, 
there is a significant deficit of massive stars in the 
low-density Orion A cloud (Hsu et al. 2013).
Indeed, the existence of a correlation between the 
most-massive-star ($m_{\rm max}$) and the stellar mass in the 
embedded cluster ($M_{\rm ecl}$) has 
been found to be highly significant 
(Weidner et al. 2013,2014; 
Stephens et al. 2017;
Ramirez Alegria et al. 2016).

\begin{figure*}
\psfig{file=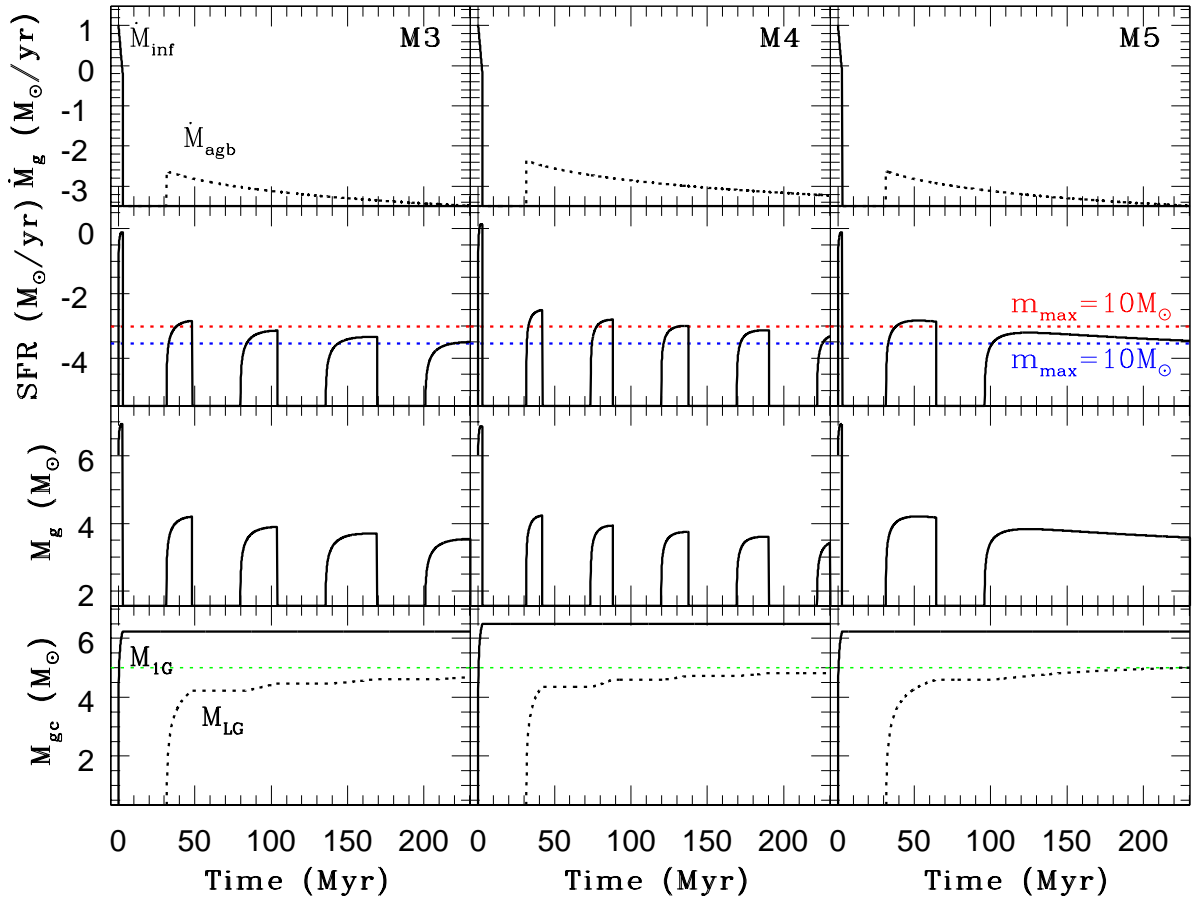,width=16.cm}
\caption{
Time evolution of
gas accretion rate and AGB ejection rate (top),
SFR (second from the top),  
total gas mass (second from the bottom) 
total masses of stars in 1G and LG (bottom),
for M3 (left), M4 (middle), and M5 (right).
The blue (upper) and red (lower) horizontal lines (in the SFR
evolution) represent the SFRs corresponding to 
 $m_{\rm max}=10 {\rm M}_{\odot}$ in model A and B, respectively.
The green line in the bottom panel indicates the typical mass
of 2G stars observed in the Galactic GCs (C09).
}
\label{Figure. 6}
\end{figure*}

\begin{figure*}
\psfig{file=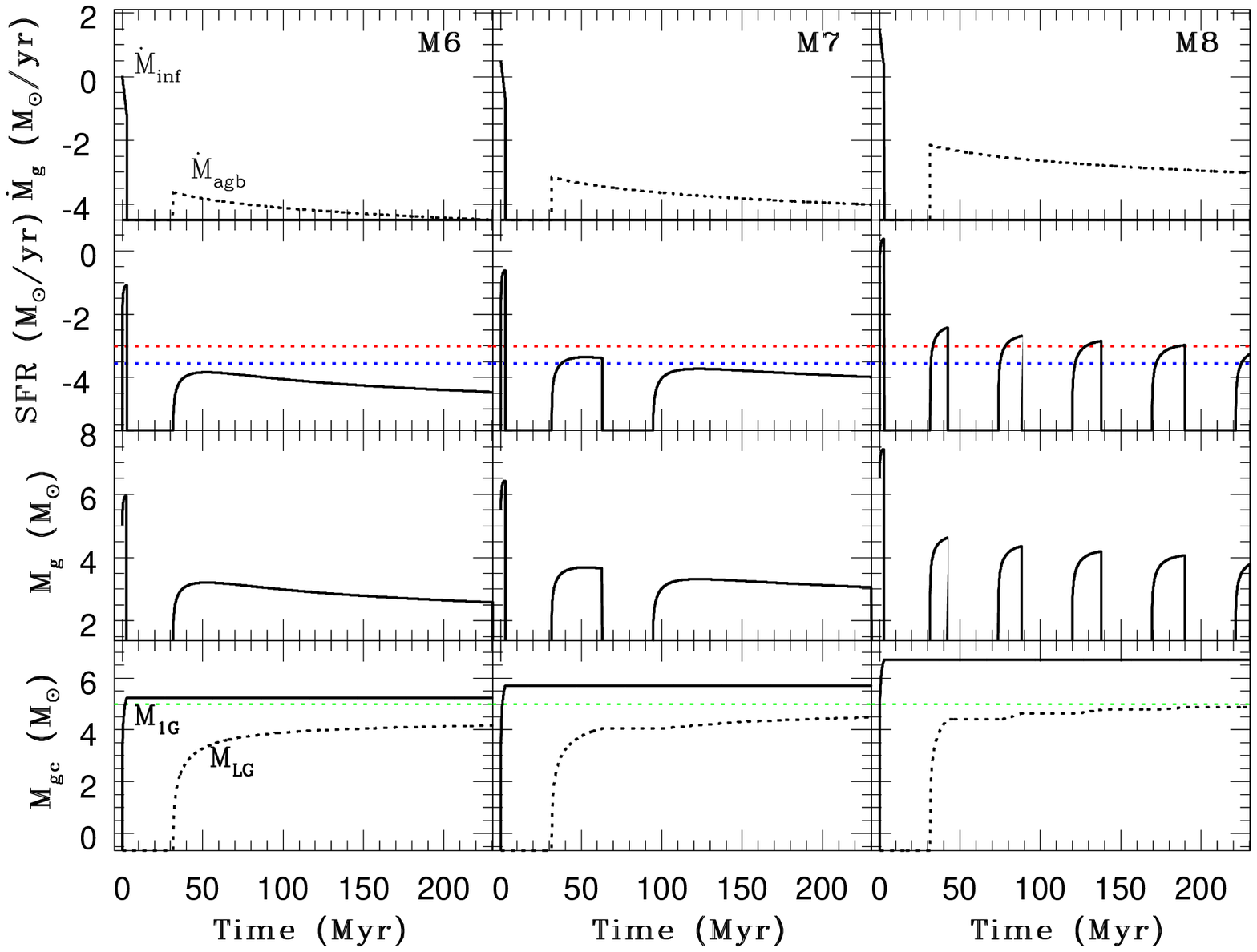,width=16.cm}
\caption{
The same as Fig. 6 but 
for M6 (left), M7 (middle), and M8 (right).
}
\label{Figure. 7}
\end{figure*}

The mathematical formalism of the IGIMF theory 
(e.g. Recchi \& Kroupa 2015; Fontanot et al. 2017; Yan et al. 2017)
can  be applied to the problem at hand, 
namely the formation of the LG from AGB ejecta in young GCs.  
The key relation which is of relevance for this 
problem is the $m_{\rm max}-SFR$ relation and as 
well that the IMF of the LG (i,e. the IGIMF) has a 
non-canonical shape above about $1\,M_\odot$ by being top-light for the 
relevant physical situation here given by the small SFRs of the 
LGs.  The combination of these two implications lead to a significantly 
smaller number of SN events produced by the LGs such that star-formation from 
AGB ejecta can, contrary to the case of an invariant canonical IMF, 
build-up to a very significant mass with the added conclusion that this 
LG star formation must be truncated semi-periodically by the few 
SN events that do occur in the LG when the SFR is sufficiently 
high. This is shown by simulations in the following sections.

\subsection{Chemical yields of AGB ejecta}

The present one-zone model is quite different from B17 in that it
does not include chemical evolution explicitly. Accordingly,
the model can not predict the chemical abundances of GC stars in a 
fully self-consistent way.  However, we investigate the possible
internal spreads in chemical abundances of light elements
among GC stars by allocating the  abundances
for each star formation episode in GC formation.  We particularly investigate
the [Mg/Fe] and [Al/Fe] abundances of GC stars by using the chemical yields
table derived by Ventura et al. (2009, 2011). 
Since the tables by Ventura et al. (2009, 2011) are only for 12
different masses of stars (i.e., 12 metallicity bins), we use an interpolation method
to calculate [Mg/Fe] and [Al/Fe], if a star has an abundance
between two metallicity bins.

We investigate the [Mg/Fe]-[Al/Fe] anti-correlation 
which is already known to be produced by the AGB scenario 
reasonably well (e.g, B07, D10,
 Ventura et al. 2016). 
The currently available chemical yields tables from Karakas 
(2010) and Ventura et al. (2011), 
however, 
cannot reproduce the Na-O anti-correlation so well.
This possibly problematic feature of the  AGB scenario in general 
is beyond the scope of this paper 
(we may focus more on this issue in forthcoming papers). 

\subsection{Parameter study}

Although we have  investigated many models with different model
parameters (e.g., $C_{\rm sf}$, $t_{\rm a}$, $\alpha$, $\Delta t_{\rm sf}$,
$M_{\rm mc}$ etc), we here describe the results of 11 representative
models. The parameter values  for these models are given in Table 2.
The model M1 is the continuous model in which the SF histories of GCs
is not 
influenced at all by SNe (which is highly unrealistic).
Other models, M2-M10, are the multiple burst models, which can show
multiple generations of stars in forming GCs.
Since the purpose of this paper is to understand the origin of
discrete multiple stellar populations of GCs, we almost exclusively
describe the results on these multiple burst models in the present
study. The results of the multiple burst models with a canonical
IMF for all stellar populations
(M9) and a top-heavy one for 1G stars (M10) are given in Appendix A,
because their results are not  as important compared with 
the other models.
Also the results of M11 are not discussed here,
because the final mass of 1G stars is less than 10\% of the original gas
mass, which means that the initial cluster can be completely disintegrated
after gas removal owing to the low star formation efficiency
(e.g., Hills 1980): see Appendix B for
the results.
The details of the IGIMF theory are also summarized in Appendix C.

\begin{figure*}
\psfig{file=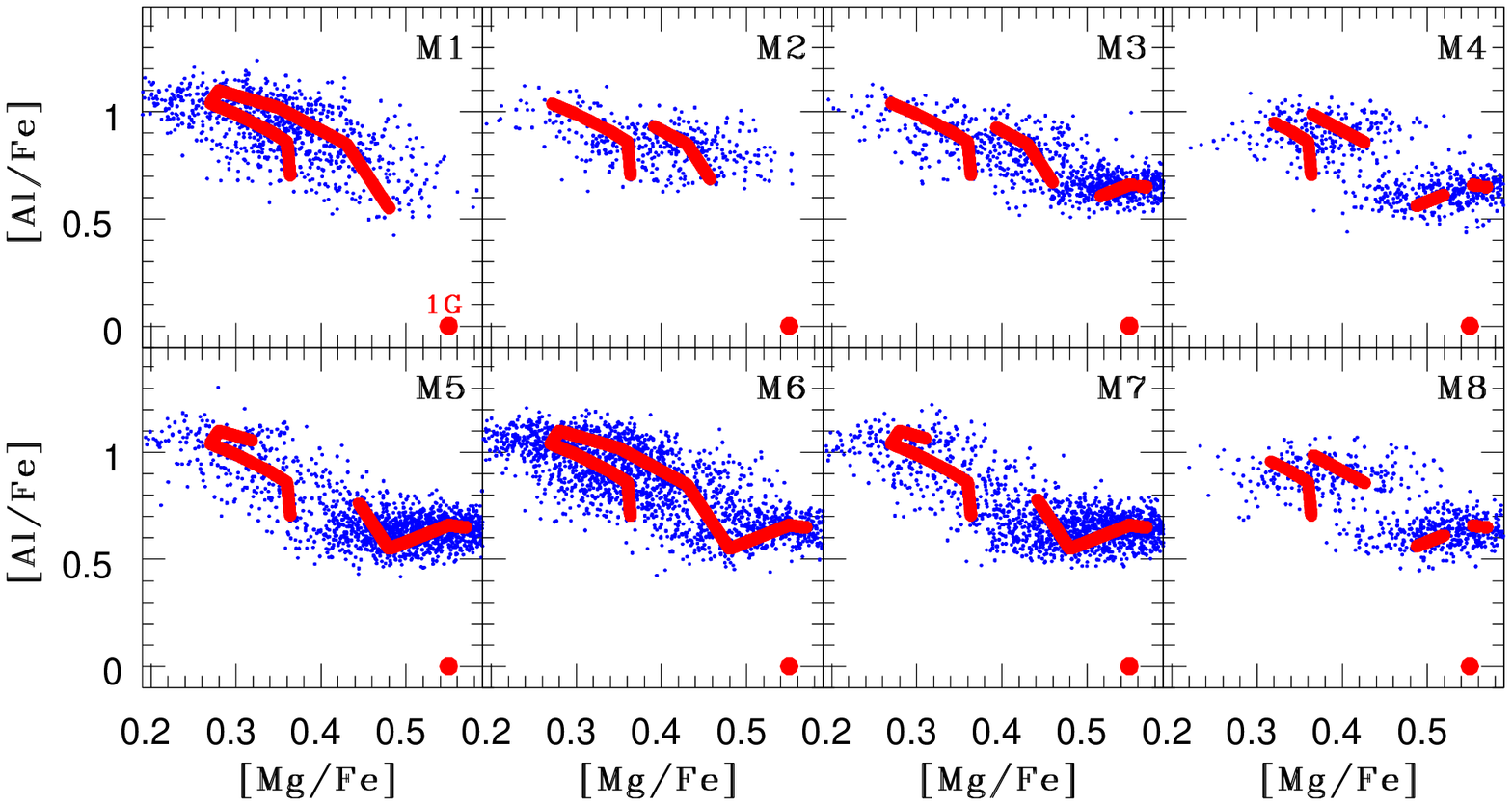,width=16.cm}
\caption{
Distributions of LG stars (big red dots) 
on the [Mg/Fe]-[Al/Fe] diagram for
the eight models (M1-M8).
The small blue dots show the distribution of GC stars 
created by adding random observational errors of 0.05 dex
to the original simulation data sets.
Each model ID is indicated in the upper right corner
of each frame. For comparison, the location of 1G stars is indicated
by one big red dot with  characters ``1G''  above it.
}
\label{Figure. 8}
\end{figure*}

\section{Results}

\subsection{Star formation histories of forming GCs}

Fig. 2 shows that 1G stars with $M_{\rm 1G}=1.7 \times 10^6 {\rm M}_{\odot}$
can be formed from a molecular cloud with $M_{\rm mc}=10^7 {\rm M}_{\odot}$
within a timescale of $\approx 3$ Myr in the continuous model M1.
The star formation rate for 1G population 
is as high as [0.1-1]${\rm M}_{\odot}$ yr$^{-1}$ owing to the initial
high gas density of the MC.
The star formation of 1G stars is truncated by the most massive SNe
with $m>100 {\rm M}_{\odot}$ in the model with a canonical IMF.
All of the remaining gas within the MC ($8.3 \times 10^6 {\rm M}_{\odot}$)
is expelled from the MC by the SNe so that the gas can not be converted
into new stars after $T = 3.2$ Myr (i.e., star formation efficiency,
$\epsilon_{\rm sf} = M_{\rm 1G}/M_{\rm mc} = 0.17$).
About $30$ Myr after the initial burst of star formation,
massive AGB stars with $m=9 {\rm M}_{\odot}$ start to inject
the gas into the 1G stellar system.
 The gas is then slowly
converted into 2G stars 
with a very low SFR ($<10^{-2} {\rm M}_{\odot}$ yr$^{-1}$)
and this secondary star formation can continue
until intermediate-mass stars with $m=4 {\rm M}_{\odot}$ enter into
their AGB phases ($T=120$ Myr). The total mass of the 2G stars can finally
become $M_{\rm 2G} \approx 10^5 {\rm M}_{\odot}$, being consistent
with the observed typical mass of 2G stars for the Galactic old GCs
(C09).

Fig. 3 shows that if we adopt the IGIMF model which depends on the SFR,
then the IMF of the GC in M1 can evolve significantly
even within 120 Myr. The number of SNe II ($N_{\rm SNII}$) predicted from the
IGIMF (181) is much smaller than that from the canonical IMF ($\approx 1100$)
for the 2G stars (Also see Fig. 2 for the evolution of $N_{\rm SNII}$).
Furthermore, $m_{\rm max}$ is smaller in the later phase of 2G formation
(e.g., $24 {\rm M}_{\odot}$) at $T=40$ Myr and $17 {\rm M}_{\odot}$
at $T=110$ Myr)
and the mean $m_{\rm max}$  is smaller 
in the IGIMF ($24 {\rm M}_{\odot}$ than in the canonical IMF 
($=120 {\rm M}{\odot}$).
These results are due to the very low SFR of 2G formation for which
the IGIMF theory predicts lower $m_{\rm max}$.
However, given the SFR,
the derived $m_{\rm max}$ from the IGIMF can not be smaller
than $10 {\rm M}_{\odot}$ in this model, which means that SNe should be able to
severely suppress or even truncate the 2G formation. 
Therefore,  the continuous formation of 2G stars over $100$ Myr
is inconsistent with $m_{\rm max}$ and a large number
of SNe in the continuous model with a canonical IMF.
We thus suggest  that previous models of GC formation
based on self-enrichment by AGB stars are  not so realistic.

Fig. 4 shows the results of M2 in which the  SF duration 
of LG stars is self-consistently derived from the adopted 
$SFR-\Delta t_{\rm sf}$  relation.
Although
star formation in the 2G formation
can be truncated by SNe in this model,
the SF duration can be as long as $16$ Myr owing
to the low-mass  of the most massive SN
with $m=m_{\rm max}=24 {\rm M}_{\odot}$.
It should be stressed here that if the canonical IMF is adopted indeed,
then the SF duration is only  $\approx 3 \times 10^6$ yr owing
to $m_{\rm max} \approx 120 {\rm M}_{\odot}$. 
After the explosion
of the lowest-mass SN ($m=9 {\rm M}_{\odot}$), 3G stars are formed
from AGB ejecta with a lower SFR. The duration of 3G formation can be 
longer than that of 2G owing to the lower $SFR$ of 3G formation
(i.e., lower $m_{\rm max}$).

Fig. 5 demonstrates that not only
the shape of the IMF but also $m_{\rm max}$ can be different
between 2G and 3G stars, if the IGIMF theory is adopted
as a basis of the calculation.
This model can not have $m_{\rm max} <10 {\rm M}_{\odot}$ 
in 2G and 3G formation,
if the standard model A for the $SFR -\Delta t_{\rm sf}$ relation
is adopted. The large number of SNe ($N_{\rm SN II}=131$)
means that the remaining AGB ejecta
can be blown away by SNe,
which is  consistent with the truncation of SF by SNe in this model M2.
The final mass of LG stars in this model, however, is 
$3.0 \times 10^4 {\rm M}_{\odot}$, which is significantly lower
than that of M1. This suggests that $M_{\rm mc}$ or $m_{\rm agb,l}$
should be larger and lower, respectively, for such multiple burst model
to have $M_{\rm LG} \approx 10^5 {\rm M}_{\odot}$.

Fig. 6 shows the three multiple burst models with 
$m_{\rm agb, l}=3 {\rm M}_{\odot}$ yet different 
$C_{\rm sf}$ and $SFR - \Delta t_{\rm sf}$ relations.
The GC in M3 can finally have 1G, 2G, 3G, and 4G with 
$M_{\rm LG}=4.8 \times 10^4 {\rm M}_{\odot}$ 
at $T=230$ Myr. 
M4 with  larger $C_{\rm sf}$ (thus larger $M_{\rm 1G}$)
shows 5 distinct generations of stars with  a higher
SF and shorter SF duration
due to the higher SFR in each SF episode. The final $M_{\rm LG}$
can be therefore larger ($6.8 \times 10^4 {\rm M}_{\odot}$) than 
that of  M2.
M6 with the model B $SFR-\Delta t_{\rm sf}$ relation
shows only two LG populations with the 3G still  forming 
at $T=230$ Myr.The very long SF duration in the 3G 
results from  the lower $m_{\rm max}$. 
This  lower $m_{\rm max}$  is due to the adopted 
$SFR-\Delta t_{\rm sf}$ relation (model B) in this model.
The GC in this model can finally have 
$M_{\rm LG}=1.0 \times 10^5 {\rm M}_{\odot}$ as M1,
which is more consistent with observations (C09).
Although there is little scatter in the observed
$SFR-m_{\rm max}$ relation 
(thus $SFR - \Delta t_{\rm sf}$ relation)
for star-forming regions of the Galaxy (Weidner et al. 2013),
the relation can be different in the central region of
forming GCs, as mentioned above.
Thus,
the results of M5 with a non-standard
$SFR - \Delta t_{\rm sf}$ relation are quite interesting in the sense
that $M_{\rm LG}$ is more consistent with the observed typical
value for the Galactic GCs than M2 with the standard relation.

Fig. 7 describes  how the SFRs of forming GCs depend on
their parent MC masses. The low-mass MC model M6
with $M_{\rm mc}=10^6 {\rm M}_{\odot}$
shows   no truncation of star formation in LG,
because $m_{\rm max}$ becomes less than $9M_{\odot}$
owing to very low SF during 2G formation.
This result implies
that low-mass GCs are unlikely  to have discrete
multiple stellar populations. 
Although the less-massive model M7 with $M_{\rm mc}=3\times 10^6 {\rm M}_{\odot}$
shows 2G and 3G formation, 
$m_{\rm max}$ becomes less than $9M_{\odot}$ in the 3G formation
so that 3G star formation can not be truncated.
Interestingly, the mass-ratios of LG to 1G in these models
are higher than those derived for M3 
with $M_{\rm mc}=10^7 {\rm M}_{\odot}$,
because the formation of LG stars can continue longer in these models.

Fig. 7 also shows that 
the SFR during the formation of LG stars is significantly higher in
the massive GC model
M8 with $M_{\rm mc}=3\times 10^7 {\rm M}_{\odot}$.
As a result of this,
the duration of the star formation is  shorter
owing to higher $m_{\rm max}$ ($\approx 60 {\rm M}_{\odot}$).
Therefore, the GC in M8 can finally have 5 distinct stellar populations
at $T=230$ Myr: one more stellar population in comparison with M2.
The shorter duration of star formation in 1G and 2G 
implies that the abundance spread in these generations
can be small.
The final mass of LG stars ($M_{\rm LG}$) becomes
as large as $10^5 {\rm M}_{\odot}$,
which is consistent with the observed typical $M_{\rm LG}$ of
the Galactic GC.
These results of the three models suggest that the numbers of discrete
stellar populations in GCs depend on the initial masses of GCs. 

\section{Discussion}

\subsection{Internal abundance spreads in discrete stellar populations}

The multiple generations of stars formed in several GC formation models
of the present study (e.g., M2 and M3) strongly suggest that
discrete stellar populations can be seen in the 
[Na/Fe]-[O/Fe] and [Mg/Fe]-[Al/Fe] diagrams
of GC stars.
It is thus our important investigation 
whether such discrete populations can be really seen
in the [Mg/Fe]-[Al/Fe] diagram.
The purpose of this investigation is not to reproduce the
observed [Mg/Fe]-[Al/Fe] anti-correlations (e.g., Ventura et al. 2016)
but to illustrate whether distinct groups of stars in
the [Mg/Fe]-[Al/Fe] diagram can be clearly seen in
the present models.
More detailed investigation of each individual objects (such as M13)
will need to be done in our future studies with more sophisticated
modeling of GC formation.
As shown in Fig. 8,
the GC at the final time step in M1 has two distinct groups of stars 
on the [Mg/Fe]-[Al/Fe] diagram, and the two groups are well separated
from the location of 1G stars. The GC in M1, on the other hand,
has a continuous distribution of stars along the expected
[Mg/Fe]-[Al/Fe] anti-correlations.
The GCs in the models M3 and M4 with  longer duration of LG star formation
has four and five discrete stellar populations, respectively,
though the locations of 3G, 4G, and 5G on
the diagram are hard to be distinguished in M3.
A wide gap between the locations of 2G and 3G stars can be seen in the
GC of M3 with a long duration of 3G star formation.

As expected from star formation histories of GCs in M6-M8,
only more massive GCs formed from 
$M_{\rm mc} \ge 3 \times 10^6 {\rm M}_{\odot}$
can have discrete stellar populations on the [Mg/Fe]-[Al/Fe]
diagram. The distribution of  each sub-population  appears to be
narrower in M8, which reflects the fact that the duration of LG star formation
episodes are 
shorter in this model. This suggests that discrete stellar populations
on the [Mg/Fe]-[Al/Fe] diagram
can be more clearly seen 
in more massive GCs.
Although this `mass-dependent visibility of discreteness'
is an important prediction of the present model of GC  formation
based on the IGIMF,
it is at present  not observationally clear whether
such discreteness can be more clearly seen in more massive GCs
owing to the small number of GCs investigated 
with more precise abundance measurements (e.g., C14).

Errors in observational measurements
of [Mg/Fe] and [Al/Fe] can broaden the narrow distribution of 
stars along the original tight Mg-Al anti-correlation
so that the discrete stellar populations may appear  much less
pronounced. 
Observational errors in the abundance estimation ([A/Fe],
where $A$ is a light element)
ranges typically from 0.031 to 0.076 in C09,
which is not negligible in comparison with
the possible differences of [A/Fe] between LG stars predicted
from the present study.
It is therefore possible that 
the observed apparent continuous
distributions of GC stars along the Mg-Al anti-correlation
is indeed due to the broadening through
observational uncertainty  of the original discrete
stellar populations (C14).
In order to investigate this issue, we created new distributions of
GC stars on the [Mg/Fe]-[Al/Fe] diagram in each model  by adding random errors
to the original simulation data. 
Dispersions of 0.03, 0.05, and 0.1  dex are
added to [Mg/Fe] and [Al/Fe] in the original
simulation data  in order for us to investigate whether original
discrete distributions of GC stars can disappear owing to the
addition of such small observational errors.

Small dots in Fig. 8 show the 
the new distributions that are created by adding 
0.05 dex random errors to the original simulation data.
Clearly the distributions  now appear much less discrete 
along the Mg-Al anti-correlation. It is furthermore
confirmed that this almost disappearance of
the discrete distributions can be seen in the new
distributions with 0.03 and 0.1 dex random errors added to
the original data.
These  results therefore
suggest that  the observed apparently continuous distributions of
GC stars along the Mg-Al anti-correlation 
(and Na-O one) do not necessarily mean that the true
distribution
is  continuous too.
They also suggest that more precise estimation of Mg and Al abundances
is necessary for many GCs
to discuss whether GCs experienced  multiple episodes of
star formation separated by of the  order of $10^7$ yr.

The key physical processes for the formation of discrete
multiple populations  are the following three.
First is the truncation of star formation from AGB ejecta
by SNe in each generation of
stars. Second is the relatively long interval ($\approx 20$ Myr)
between the formation of two generations of stars. 
Third is  the time-evolving IMF during GC formation, which
ensures the longer duration ($>10$ Myr) 
of star formation in each generation.
Although the `top-light' IMF is required to lengthen
the duration of star formation,
an overly top-light IMF without  SNe can not explain
the presence of discrete stellar populations in GCs,
because it allows AGB ejecta to continue to be converted into new stars.
As demonstrated in Appendix A,  the model with a  canonical universal IMF 
(M9) predicts very small total  masses  of later generations of stars
($M_{\rm LG}=0.0002M_{\rm mc}$), which means that typical GCs
would need to be formed from super-giant MCs 
with $M_{\rm mc} \approx 5 \times 10^8 {\rm M}_{\odot}$.
But this required $M_{\rm mc}$ appears to be too large to be realistic:
GC formation models with a time-evolving IMF appears to be  more realistic.

The present study predicts that more massive GCs are more likely to
have at least a few  discrete multiple  stellar populations. 
Such massive GCs can possibly have six discrete populations, 
if new stars can continue to 
be formed from ejecta of AGB stars until stars with $m=3 {\rm M}_{\odot}$
become AGB stars.
(i.e., until $\approx 400$ Myr after the initial starbursts).
So far NGC 6752, M22, and NGC 2808 have been observed to show three or more
discrete stellar populations in the [Na/Fe]-[O/Fe] and
[Mg/Fe]-[Al/Fe] relations  and the color-magnitude
diagrams (C14; Marino et al. 2011) and these are massive GCs.
Currently it is not observationally clear whether GCs with lower masses
(yet with multiple stellar populations) have discrete stellar populations
or not.  Apparently, all of the GCs with multiple stellar populations
in C09 appear to show continuous populations in the [Na/Fe]-[O/Fe]
diagrams. However, these apparently continuous stellar populations
could be due largely to observational errors  in the estimation of
chemical abundances by previous spectroscopic observations.
It is thus doubtlessly worth while for observational studies
to investigate the distributions of stars on the 
[Na/Fe]-[O/Fe] and  [Mg/Fe]-[Al/Fe] diagrams of many GCs
to gain a  better understanding
of the origin of the discrete multiple stellar populations.

\subsection{Lack of OB stars in LG star formation $-$ the case
of young massive clusters (YMCs) }

Recent observational studies have searched for evidence of ongoing
star formation in young massive clusters (YMCs) in nearby galaxies
and found no evidence for it (e.g., Larsen et al. 2011;
Bastian et al. 2013;
Cabrera-Ziri et al. 2014). For example,
Bastian et al. (2013) investigated H$\beta$ and [OIII] emission lines
of 130 YMCs with ages ranging from 10 Myr to 1000 Myr and found no evidence
of such emission lines in the YMCs. Since stellar radiation from massive
OB stars are responsible for such emission lines,   
they concluded that secondary star formation lasting
over hundreds of Myr can be ruled out by their observations.
Cabrera-Ziri et al. (2014) 
investigated the stellar population of a YMC  with a mass
of $\approx 10^7 {\rm M}_{\odot}$ and found that its 
spectral energy distribution (SED) is consistent
with a single stellar population with an age of $\approx 100$ Myr.
Goudfrooij et al. (2014), however, criticized these interpretations
of YMCs and pointed out that the observations can not rule out
the secondary star formation in the YMCs.

The present study has shown that 
$m_{\rm max}$ (the maximum mass of stars) can be well less than
$30 {\rm M}_{\odot}$ during secondary star formation in star clusters.
Therefore,  the observed apparent lack
of massive OB stars in YMCs of
nearby galaxies can result not from  the absence of
secondary star formation within the YMCs but from a
lower $m_{\rm max}$ in secondary star formation within the YMCs.
Accordingly, the lack of H$\alpha$ emission lines in  YMCs
can not rule out secondary star formation: other diagnostic observations
are required to distinguish between the above two scenarios for
the lack of H$\alpha$ emissions in clusters.
If LG stars are formed with a top-light IMF (i.e., low $m_{\rm max}$) in 
a forming GC,
as demonstrated in the present study,
then the  SED of the GC would not be influenced by
the less dominant LG stars when the GC is $\approx 100$ Myr old.
That star formation can proceed without the formation of massive stars has been
documented already in several observational papers
(e.g., Kirk \& Myers 2011; Hsu et al. 2013).

Indeed, For \& Bekki (2017) have recently discovered young stellar objects
(YSOs) within older star clusters with ages ranging from 0.1 Gyr to
1 Gyr in the LMC using the observational  data
obtained by Spitzer and Herschel.
This discovery  has clearly 
demonstrated ongoing (secondary) star formation  is possible
in the older star clusters of the LMC and would be possible
in other YMCs of other galaxies.
This discovery by large infrared
telescopes suggests that secondary star formation can be more easily detected
in infrared observations
than in optical photometric and spectroscopic ones.
It is currently a formidable task for observational studies to
find dust-shrouded intermediate-mass and low mass stars with ages
less than 1 Myr in YMCs of nearby galaxies. 
We suggest, however,  that future observational studies on
the presence or absence of YSOs in YMCs are indispensable for
proving if secondary star formation in older YMCs can be ongoing.

\subsection{Dilution of AGB ejecta with pristine gas}

So far we did not discuss the importance of dilution of AGB ejecta
with ``pristine gas'' (i.e., gas that has the same chemical abundances
as 1G stars) in the formation of multiple stellar populations of GCs.
Previous one-zone chemical evolution models of GC formation
already demonstrated that such dilution processes are essential for
reproducing the observed Na-O, Mg-Al, and C-N 
anti-correlations and He abundance distributions
(B07, D10). 
Accretion of gas onto the 
potential of a cluster from the surrounding ISM was 
expected from analytical calculations (Pflamm-Altenburg \& Kroupa 2009).
Recent hydrodynamical simulations of GC formation have demonstrated
that gas accretion onto 1G stars of a GC  from cold ISM surrounding the GC is
possible in a GC-host dwarf galaxy (B17a). Such gas accretion has been
demonstrated to occur $\sim 50$ Myr after the commencement of accretion
of AGB ejecta onto the GC (Fig. 8 in B17a). This implies that new (LG) stars
can be formed from pure AGB ejecta, if the new star formation
occurs less than 50 Myr after  gas ejection
from the most massive AGB stars ($ {\rm M}_{\odot}$).
A very  small  fraction ($\approx 1$\%)  of cold gas within GC-forming MCs
cannot be influenced by SNII and thus
can be re-accreted onto 1G stellar systems (Fig. 1 in B17b).
Such cold gas can be mixed with AGB ejecta to form LG stars (B17b).
Gratton \& Carretta (2010) proposed that gas from unevolved stars
can be used for dilution of AGB ejecta.

Although the above-mentioned previous works clearly suggested the importance
of dilution of AGB ejecta in GC formation,
the time evolution of gas accretion onto existing
GCs cannot be simply approximated by an analytic function. Therefore, we here
discuss how the dilution process can possibly change the present results by
assuming that AGB ejecta is mixed with the same amount of pristine gas: this
assumption on the amount of pristine gas is quite reasonable (B07, D10).
A factor of two increase in SFR 
is expected in this model with dilution owing to the adopted star-formation model.
The adopted $\Delta t_{\rm sf}- \log {\rm SFR}$
relation for Model A (in Fig. 1) can be approximated
by the following linear relation for $\log {\rm SFR} \le -3$ 
(${\rm M}_{\odot}$ yr$^{-1}$):
\begin{equation}
\Delta t_{\rm sf}= -13.3 \times (\log {\rm SFR}+3.5)+25.3.
\end{equation}
This relation means that a factor of two increase in SFR from
$\log {\rm SFR}=-3$ to
$\log {\rm SFR}=-2.7$ 
(due to dilution)
can shorten the duration of LG formation from 27.9 Myr to 21.8 Myr
($\sim 22$\% decrease). 
This level of shortening is not so significant, and accordingly it would not 
change the present results (e.g., the total mass of LG stars etc) significantly.
Therefore, we conclude that the roles of IGIMF in the formation
of discrete multiple populations 
do not depend strongly on whether dilution is included in the models or not.

Milone et al. (2017) 
 have recently characterized the multiple stellar populations
of 57 GCs using the photometric data from {\it Hubble Space Telescope (HST)}
UV Legacy Survey of Galactic Globular Clusters. One of their remarkable results
relevant to the present study is that some GCs show the split
of both 1G and 2G populations (``Type II cluster'' in their definition).
They have shown that
these Type II GCs with (at least) four populations are also enriched
with iron and $s$-process elements.
They have also found that  some GCs have distinct stellar clumps
along the 1G and 2G sequences whereas others have no such clumps.
Although the presence (or absence)
of such distinct groups
of stars can be more robustly confirmed
using a large number of stars (Milone et al. 2017),
the origin of these GCs with distinct multiple 
stellar populations can be explained by the scenario presented by this study.

The apparently smooth distributions along the 1G and 2G sequences observed
in some GCs (Milone et al. 2017) cannot be simply explained by the present multiple-burst
models. If such smooth distributions are observed only for low-mass GCs in 
Milone et al. (2017),
they are consistent with the prediction of the present study
(see Fig. 7).  However, it is not clear whether GCs with
such smooth distributions are more likely to be low-mass GCs in
 Milone et al. (2017).
As shown in previous one-zone chemical evolution models with
continuous star formation (B07 and D10),
mixing (i.e., dilution)
of AGB ejecta with pristine gas and the subsequent star formation
would  better explain the origin of such GCs.
It is not well understood, however, how such dilution is possible
in the early formation histories of GCs.

\section{Conclusions}

The present study has adopted a new GC formation model with a time-varying
IMF (based on the IGIMF theory)
in order to discuss the origin of discrete multiple stellar populations
in GCs. The new model has incorporated the  $SFR$-$m_{\rm max}$
relation (e.g., Yan et al. 2017) to estimate the duration of star formation in 
forming GCs in a self-consistent manner.
For comparison, we have also investigated models with a constant 
universal IMF. We have adopted a reasonable assumption
that new stars can be  formed from AGB ejecta which accumulates in
the young  GC.
The basic parameter in this study is the initial mass of the GC-forming
molecular cloud (MC; $M_{\rm mc}$) and the lower cut-off mass of AGB stars
($m_{\rm agb, l}$) above which AGB ejecta can be used for secondary star
formation.
The principal results are as follows: \\

(1) The second generation (2G) of stars can be formed from
AGB ejecta of the first generation (1G) of stars $\approx 30$ Myr
after the initial starburst of 1G stars within a GC
in the fiducial model with $M_{\rm mc}=10^7 {\rm M}_{\odot}$
and a non-universal IMF. However,
the 2G star formation is truncated by SNe $\approx 16$Myr after
the commencement of 2G star formation. This longer
duration of 2G star formation
results from  the longer lifetimes of the most massive
stars (i.e., lower $m_{\rm max}$ $\approx 24 {\rm M}_{\odot}$)
in the model.
This lower $m_{\rm max}$ is due to the
low SFR ($<10^{-2} {\rm M}_{\odot}$ yr$^{-1}$) in the 2G
star formation from AGB ejecta. Star formation is inhibited for
about 30 Myr until the last 2G SNII explodes.

(2) The third generation (3G) of stars are then formed from
AGB ejecta of 1G stars $\approx 30$ Myr after the truncation of 2G star
formation in the fiducial model.
This cycle of the formation of new stars  and the abrupt
truncation of star formation by SNe from the new massive stars
in a forming GC
can continue until the GC  loses the intra-cluster gas by some physical
processes, such as ram pressure stripping of the gas by the Galactic hot
halo gas (e.g., Frank \& Gisler 1976; Bekki 2006).
The duration of star formation is longer for later generations of stars,
because SFRs are  lower (i.e., $m_{\rm max}$ is lower) in the later
generations such that the SNe explode after longer time scales.
Thus, it is inevitable that forming GCs experience a number of star 
formation episodes each of which is  separated by $\approx 30$ Myr intervals.
Since chemical abundances of AGB ejecta depends on the masses of AGB stars,
new stars formed from AGB ejecta  can have different 
chemical abundances. \\

(3) The models with a constant canonical IMF also show multiple generations
of stars during GC formation.  However, the duration of each star formation
episode is too short ($\approx 3 \times 10^6$ yr)
because of the large number of SNe so that the total mass
of these later generations of stars ($M_{\rm LG}$)
 in a GC is quite small. For example,
the total masses of 1G stars 
($M_{\rm 1G}$)  and all other generations of stars 
(e.g., 2G, 3G, and 4G)
are $1.7 \times 10^6 {\rm M}_{\odot}$
 and $2 \times 10^3 {\rm M}_{\odot}$, respectively,
in the model with $M_{\rm mc}=10^7 {\rm M}_{\odot}$.
This result suggests that (i) $M_{\rm mc}$ of a GC-forming MC
should be as large as
$5 \times 10^8 {\rm M}_{\odot}$ for the GC to have 
$M_{\rm LG}=10^5 {\rm M}_{\odot}$ as observed for typical Galactic
GCs with multiple stellar populations (e.g., C09)
and (ii) almost all of the 1G stars
need to be lost for the GC to have a high ratio of $M_{\rm LG}$ to
$M_{\rm 1G}$.
Since these two requirements appear to be very hard to be met,
GC formation models with a universal IMF can be possibly ruled out. \\

(4) The present study predicts that low-mass GCs are unlikely to
have {\it discrete} multiple stellar populations, even though they
have 1G and 2G stellar populations. This is because
$m_{\rm max}$ during secondary star formation is lower
than $9 {\rm M}_{\odot}$ so that AGB ejecta can continue to be
converted into new stars without being blown away by SNe.
Accordingly, Na-O and Mg-Al anti-correlations between
stars in these low-mass GCs 
should be continuous rather than discrete.
We thus predict that more massive GCs are more likely to have
discrete multiple stellar populations. This prediction will be
assessed by future observations of chemical abundances of 
GCs with high-precision spectrograph.

(5) The present study suggests that  massive 
star clusters with ages older than $10^7$ yr
can have no OB stars, even if
they are currently forming  new stars from 
AGB ejecta. This is mainly because $m_{\rm max}$ is predicted to 
be  significantly
lower than $30 {\rm M}_{\odot}$ owing to low SFRs in these clusters.
Accordingly, the lack of OB stars in massive clusters with
ages older than $10^7$ yr does not
necessarily mean the lack of secondary star formation in the clusters.
Although massive clusters with secondary
star formation only in low-mass and intermediate-mass
stars can not show H$\alpha$ emission from OB stars,
young stellar objects (YSOs with ages less than 1 Myr)  or
gas and dust (with the total masses being 
less than $10^4 {\rm M}_{\odot}$) can be evidence for secondary
star formation in these clusters.
It is our future study to investigate whether LG stars can have different
[Fe/H] from 1G stars in the present multiple burst scenario.

\section{Acknowledgment}
We are   grateful to the referee  for  constructive and
useful comments that improved this paper.
KB, PK, and TJ  acknowledge financial support through the  DAAD
(The Australia-Germany Joint Research Co-operation Scheme)
throughout the course of this work.
TJ was support by University of Bonn and by Charles University 
through a grant SVV-260441.

\appendix

\section{Canonical and top-heavy IMFs with multiple episodes of star
formation}

As discussed briefly in the main text,
the duration of 2G star formation should be as short as $3 \times 10^6$ yr
corresponding to the lifetime of the most massive star formed during the
2G formation in M1, if a canonical IMF is adopted.
This means that the adopted continuous formation of SG stars without
being truncated by SNe is highly unrealistic.
It is thus useful for the present study to investigate how
the results of M1 can change if the truncation of star formation
by SNe is self-consistently included for a canonical IMF.
Here, star formation of LG populations (e.g., 2G, 3G etc)
is truncated by the most massive 
SNe ($\approx 100 M_{\odot}$)
only $\approx 3$ Myr after the start of the star formation
in M9. The short duration of star formation in M9
is due to the adopted assumption of a canonical IMF
with a fixed large upper mass cut-off.

Fig. A1 shows that the formation
of discrete multiple stellar populations (1G-7G) is possible
owing to the rapid cycle of star formation and its truncation by SNe.
However, the duration of 1G-7G formation is so short that the total
mass of new stars formed in each star formation episode is rather
small ($<10^3 {\rm M}_{\odot}$).
Consequently, the final $M_{\rm LG}$ is only 
$2.0 \times 10^3 {\rm M}_{\odot}$, which is by almost three orders of
magnitude smaller than $M_{\rm 1G}$.
This means that 99.9\% of 1G stars need to be preferentially lost
in the later dynamical evolution of the GC for $M_{\rm LG}$
to be comparable to $M_{\rm 1G}$.
This is the most serious version
of the classic `mass-budget' problem, which appears to be very hard to be
solved.

If the IMF of 1G stars in a forming GC is top-heavy
(e.g., Marks et al. 2012),
then the total amount of AGB ejecta can be increased to some extent in
the later formation phase of the GC.
This can cause a difference in the later formation of stars from
AGB ejecta.
Fig. 10 describes the evolution of M10 in which the model parameters
are exactly the same as those in M3 except for the IMF in the 1G formation
($\alpha=1.35$ for 1G formation).
The results are not so different from M3, which implies 
that the IMF slope in 1G formation
is not so important in the formation of LG stars
if the non-universal IMF is adopted for the LG formation.
The top-heavy IMF  can dramatically reduce the number fraction of low-mass
long-lived stars ($0.1 \le \frac{m}{ {\rm M}_{\odot} } \le 0.8$).
Therefore, the mass fraction of 2G low-mass stars to 1G ones can be
significantly larger in M10 than in M3.

\begin{figure}
\psfig{file=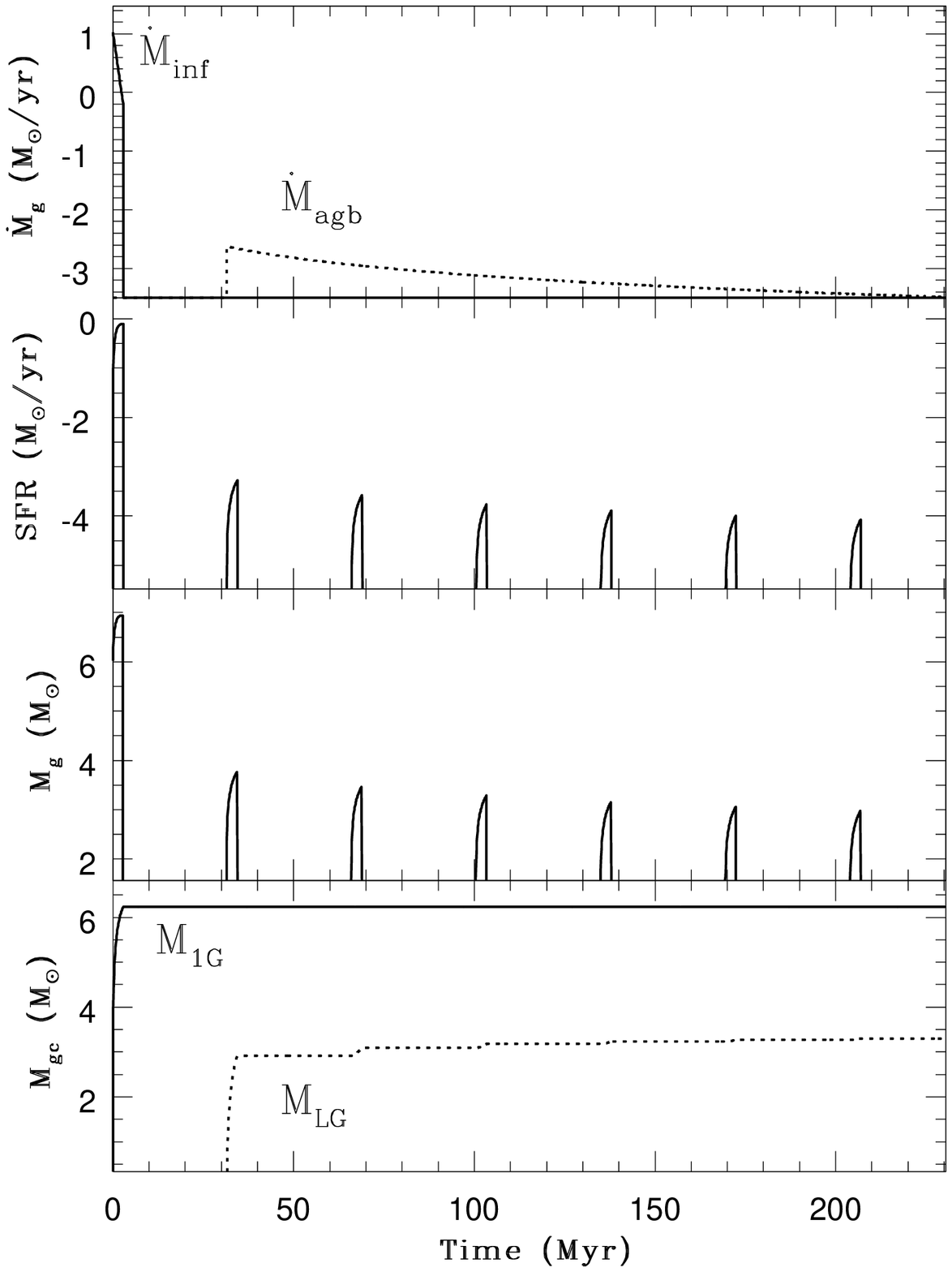,width=8.5cm}
\caption{
Time evolution of
gas accretion (${\dot{M}}_{\rm inf}$)  and AGB ejection rates 
(${\dot{M}}_{\rm agb}$; top),
SFR (second from top),  total masses of stars in 1G and LG 
(second from bottom),
and total gas mass (bottom)  in M9.
}
\label{Figure. 9}
\end{figure}

\begin{figure}
\psfig{file=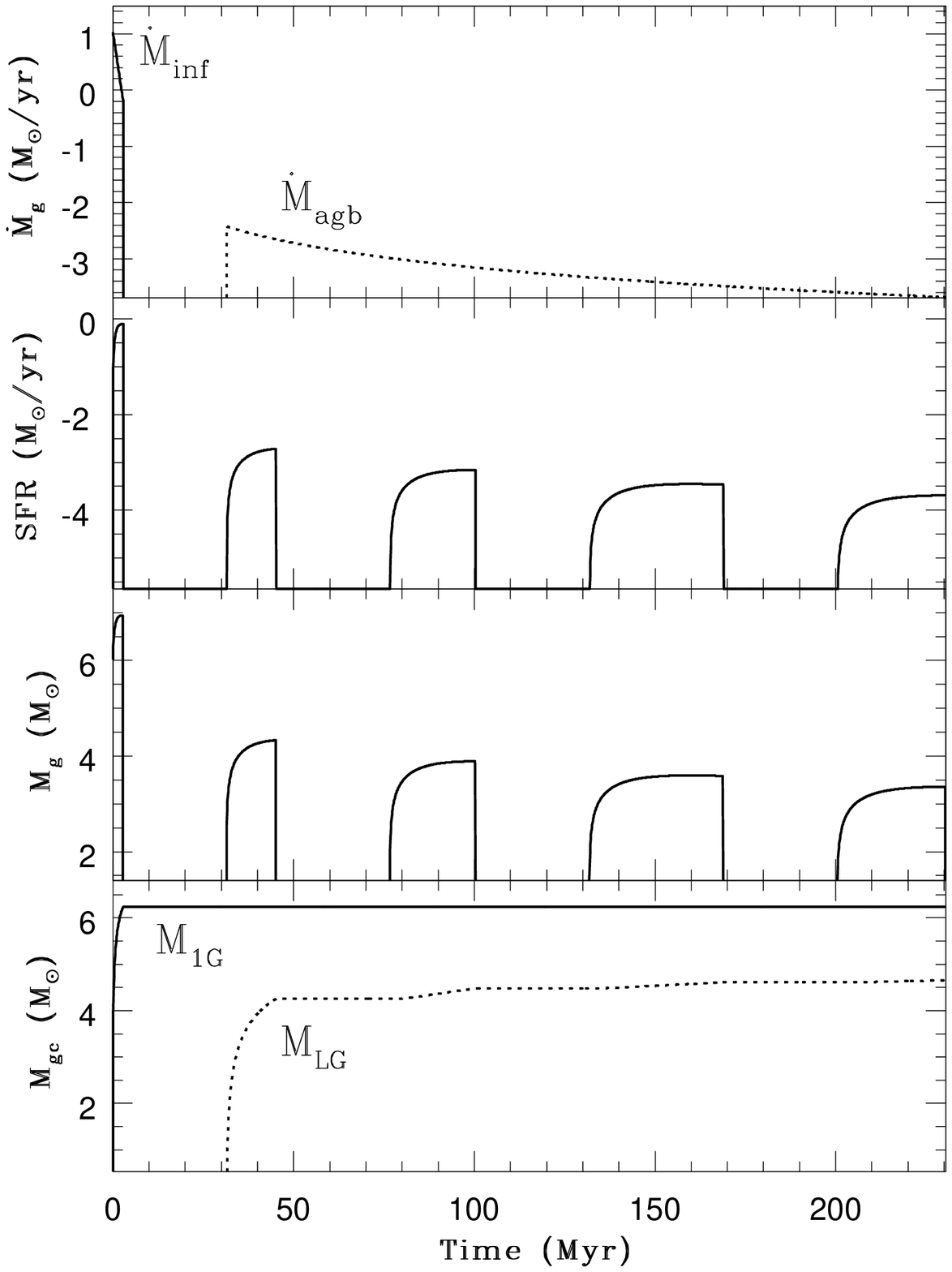,width=8.5cm}
\caption{
The same as A1 but for M10.
}
\label{Figure. 10}
\end{figure}

\section{Dependence on $C_{\rm sf}$ }

Fig. B1 shows that the time evolution of SFR, $M_{\rm g}$,
and $M_{\rm gc}$ in M11 with a smaller $C_{\rm sf}$ (=4.5)
is similar to that described for M2.
However, the final mass of 1G stars in this model is
only $9.0 \times 10^5 {\rm M}_{\odot}$, which is less than 
10\% of the original gas mass. Therefore, the 1G stellar system
is highly unlikely to be gravitationally bound after the removal of
the left-over gas (e.g., Hills 1980). It is confirmed
that the models with $C_{\rm sf} \le 4.5$
show such a low $M_{\rm 1G}$, which means  that $C_{\rm sf} \le 4.5$ is not
appropriate for a model of GC formation. 
The final total mass of LG stars in this model is also small 
($M_{\rm LG}=1.7 \times 10^4 {\rm M}_{\odot}$) owing to the smaller $M_{\rm 1G}$.
We conclude that the adopted $C_{\rm sf}=[9-18]$ is quite reasonable,
not only because the star formation histories of 1G and LG in the models
are consistent with those derived in hydrodynamical simulations of GC
formation (B17b), but also because the formation of bound clusters is possible
in the models.

\begin{figure}
\psfig{file=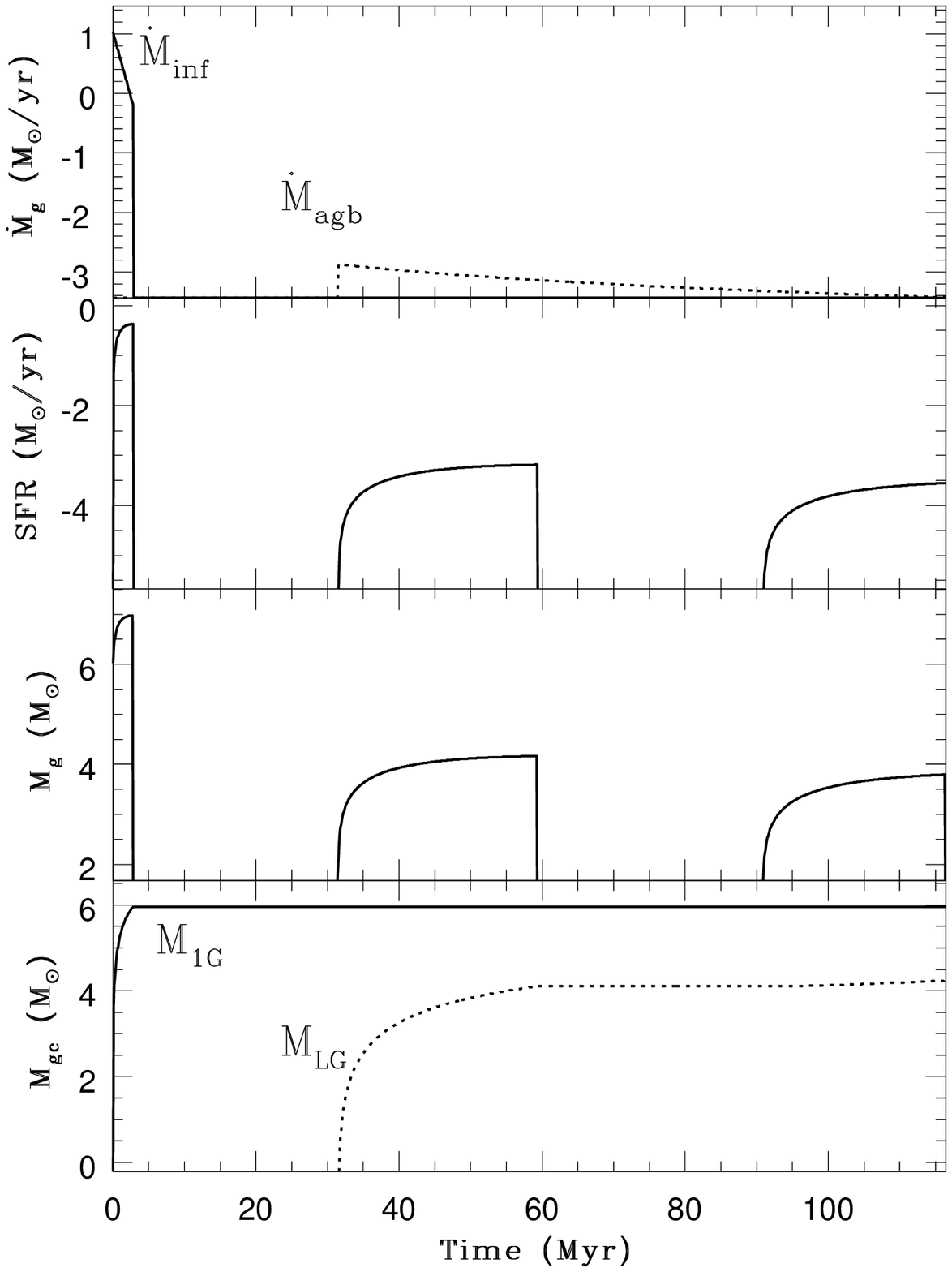,width=8.5cm}
\caption{
The same as A1 but for M11.
}
\label{Figure. 11}
\end{figure}

\section{Application of the IGIMF theory to GC formation}

The simultaneous occurrence of a largely invariant 
IMF and a $m_{\rm max}-M_{\rm ecl}$ relation is 
most likely due to feedback self-regulation of the 
process of star formation on the molecular-cloud core scale (
i.e., during the formation of an embedded cluster on a 
$<1\,$pc spatial scale and $<1\,$Myr time-scale, Kroupa et al. 2013). 
Remarkably, the form of this relation follows 
readily by a simple integration over the canonical 
IMF (e.g. Yan, Je\v{r}\'abkov\'a \& Kroupa 2017).

On the scale of whole closed star-forming systems 
(where “closed star-forming system” is one in
which the star formation occurs within 
a self-regulated potential, a molecular cloud within a 
galaxy not being a closed star-forming system)
such as galaxies, observations have shown the ensemble 
of freshly formed clusters to have a significant correlation between the 
mass in stars of the most massive embedded 
cluster and the system-wide SFR (Weidner et al. 2004;
Randriamanakoto et al. 2013). 
This $M_{\rm ecl, max}-SFR$ relation 
readily follows from an integral over the embedded 
cluster mass function (ECMF) subject to the mass being formed 
within about $10\,$Myr being the total mass in stars formed in the 
ensemble of ECs (e.g. Yan et al. 2017). Different self-regulated 
closed systems may form different ECMFs and the canonical assumption is 
that the power-law index of the ECMF, $\beta\approx2$, 
i.e. essentially being the Salpeter index. The combination of 
the $m_{\rm max}-M_{\rm ecl}$ and the $M_{\rm ecl, max}-SFR$ 
relations yields a $m_{\rm max}-SFR$ relation for the closed system 
(Yan et al. 2017): see also our Fig. 1.

In the context of the LG problem of GCs, the one-zone 
simulations presented here demonstrate that star-formation from 
AGB ejecta begins at a low level, because initially the molecular clouds 
need time to form and to buildup near the centre of the young 
GCs in which all SN events from the 1G have exploded. Thus, during the 
first about $1\,$Myr the embedded clusters will be low mass with a 
small $m_{\rm max}$. Their feeble feedback will cause the ISM 
which forms from the AGB ejecta to redistribute itself causing the 
formation of new embedded clusters within the inner region of the young GC. If the SFR of the closed system (the young GC) becomes sufficiently 
high, some massive stars may form and these may more strongly 
regulate further star formation. But over the time of about $10\,$Myr 
one may expect an ensemble of ECs to form, the most massive member of 
which will be limited by the SFR (i.e. by the amount of  
ISM mass available), just as in a galaxy.

It is emphasized here that the application of the IGIMF theory, 
as formulated elsewhere (e.g. Recchi \& Kroupa 2015; Fontanot et al. 2017; 
Yan et al. 2017), has not been adjusted to produce a wanted result here, 
and that it is based purely on empirically deduced correlations.  
Also, application of the IGIMF theory to the scale of a young GC is 
merely an assumption made here which needs further testing. 
The assumption is thus that the IMF of the LG stars in not canonical, 
but the non-canonical IMF is not adjusted arbitrarily but is defined by 
independently obtained empirical findings.

\end{document}